# IDS IN TELECOMMUNICATION NETWORK USING PCA

Mohamed Faisal Elrawy[1], T. K. Abdelhamid[2] and A. M. Mohamed[3]

[1]Faculty of engineering, MUST University, 6th Of October, Egypt
`eng_faisal1989@yahoo.com`
[2,3]Faculty of engineering, Assuit University, Assuit, Egypt
[2]`tarik_k@aun.edu.eg`, [3]`afm@aun.edu.eg`

## ABSTRACT

*Data Security has become a very serious part of any organizational information system. Internet threats have become more intelligent so it can deceive the basic security solutions such as firewalls and antivirus scanners. To enhance the overall security of the network an additional security layer such as intrusion detection system (IDS) has to be added. The anomaly detection IDS is a type of IDS that can differentiate between normal and abnormal in the data monitored. This paper proposes two types of IDS, one of them can be used as a network intrusion detection system (NIDS) with overall success (0.9161) and high detection rate (0.9288) and the other type can also be used as a host intrusion detection system (HIDS) with overall success (0.8493) and very high detection rate (0.9628) using NSL-KDD data set.*

## KEY WORDS

*IDS, NIDS, HIDS, data mining, anomaly detection.*

## 1.INTRODUCTION

In the age of information technology revolution the telecommunications networks have been developed from circuit switched network to packet switched network, after that it has Mutations enormous towards all-IP based networks. These developments make the communication of applications and services such as data and voice are being transferred on top of the IP-protocol [1].

The development of data transmission speeds in both uplink and downlink has increased considerably from the second generation (2G) of radio access networks to the third generation (3G) of radio access networks and the development of devices that subscribers of telecommunications networks make the boundary between computers and mobile phones has become unspecified.

With the smart phones, the subscriber can do almost everything and can dispense on the basic personal computers. This means that the full data on the Internet is now in the hands of each smart phone owners. Technologies in communications networks have become more progress and it has raised new unwanted possibilities. Risks and threats that were applicable only in the fixed networks are now feasible in the radio access networks. The security systems have to become more intelligent because of threats are becoming more advanced.

The basic security measurements such as firewalls and antivirus scanners cannot keep pace with the overgrowing number of intelligent attacks from the Internet. A solution to enhance the overall security of the networks is to add an additional security layer to increase the security layers by using intrusion detection systems (IDS). Intrusion Detection System (IDS) designed to complement other security measures based on attack prevention [2]. Amparo Alonso-Betanzos et al. [3] say 'The aim of the IDS is to inform the system administrator of any suspicious activities and to recommend specific actions to prevent or stop the intrusion'.

There are two types of intrusion detection, one of them is signature- based and the other is anomaly-based intrusion detection. The signature-based or misuse detection method use patterns of well-known attacks to identify intrusions [4].

The anomaly-based intrusion detection uses network traffic which has been monitored and compared versus any deviation from the established normal usage patterns to determine whether the current state of the network is anomalous. An anomalous traffic can considered as intrusion attempt.

Misuse detection uses well-defined patterns known as signatures of the attacks. Anomaly-based detection builds a normal profile and anomalous traffic detected when the deviation from the normal model reaches a preset threshold level [5].

The anomaly-based intrusion detection depends on features selection. Well selection of features will maintain accuracy of the detection while speeding up its calculations. Therefore, any reduction in the number of features used for the detection will improve the overall performance of the IDS. If there are no useless features, focus on the most important ones expected to improve the execution speed of IDS.

This increase in the detection speed will not affect accuracy of the detection in a significant way. Incorrect selection of the features may reduce the speed of the operation and reduce detection accuracy [6].

This aim of this paper is to improve the intrusion detection system by using Principal Component Analysis as a dimension reduction technique. The Paper Compares between two different features selections, i.e.6 features and 10 features. One of this features selections can be used in Network Intrusion Detection System (NIDS) and the other can be used in Host Intrusion Detection System (HIDS).

## 2. RELATIVE WORK

Chakraborty [7] has reported that the existence of irrelevant and redundant features generally affects the performance of machine learning part of the work. Chakraborty proved that good selection of the feature set results in better classification performance.

A. H. Sung et al. [8] have demonstrated that the elimination of these unimportant and irrelevant features did not reduce the performance of the IDS.

Chebrolu et al. [9] reported that an important advantage of combining redundant and complementary classifiers is to increase accuracy and better overall generalization. Chebrolu et al. [9] have also identified important input features in building IDS that are computationally efficient and effective. This work shows the performance of three feature selection algorithms: (1) Bayesian networks, (2) Classification and Regression Trees and (3) an ensemble of Bayesian networks and Classification and Regression Trees.

Sung and Mukkamala [8], have explored SVM and Neural Networks that can categorize features with respect to their importance. Use SVM and Neural Networks to detect specific kinds of attacks such as probing, DoS, Remote to Local, and User to Root. Prove that the elimination of less importance and irrelevant features has no effect on reducing the performance of the IDS.

Chebrolu et al. [9] suggested CART-BN approach, where CART has a better performance for Normal, Probe and U2R and the ensemble approach worked has a better performance for R2L and DoS. Meanwhile, A. Abraham et al. [10] proved that ensemble of Decision Tree was suitable for Normal, LGP for Probe, DoS and R2L and Fuzzy classifier was good for R2L attacks.

A. Abraham et al. [11] prove the ability of their suggested on Ensemble structure in modelling lightweight distributed IDS.

Manasi Gyanchandani et al. [12] improved the performance of C4.5 classifier over NSL-KDD dataset using different classifier combinations techniques such as bagging, boosting and stacking. Gholam Reza Zargar et al. [2] show that dimension reduction and identification of effective network features for category-based selection can reduce the processing time in an intrusion detection system while maintaining the detection accuracy within an acceptable range.

## 3. MULTIVARIATE STATISTICAL ANALYSIS

### 3.1 Distance

Many multivariate techniques applied to the anomaly detection problem are based upon the concept of distances. The most familiar distance metric is the Euclidean or straight-line distance. In most cases, it is used as a measure of similarity in the nearest neighbour method. Let $x = (x_1, x_2, x_3, \ldots, x_p)'$ and $y = (y_1, y_2, y_3, \ldots, y_p)'$ be two p-dimensional observations, the Euclidean distance between x and y is

$$d^2(x, y) = (x - y)\,(x - y) \tag{1}$$

Since each feature contributes equally to the calculation of the Euclidean distance, this distance is undesirable when different features measured on different scales or the features have very different variability. The effect of the features that have high variability or large scales of measurement would control others that have less variability or smaller scales. As an alternative, a measure of variability can be incorporated into the distance metric directly. One of these metrics is the well-known Mahalanobis distance

$$d^2(x, y) = (x - y)\,S^{-1}\,(x - y) \tag{2}$$

Where S is the sample covariance matrix.

### 3.2 Principal Component Analysis (PCA)

Naturally in intrusion detection problems Data found in high dimensions. To easily explore the data and further analysis, the dimensionality of the data must be reduced. The PCA is often used for this purpose. PCA is a predominant linear dimensionality reduction technique, and it has been widely applied to datasets in many different scientific domains [13].

PCA is concerned with explaining the variance covariance structure of a set of variables through a few new variables, which are linear combinations of the original variables. Principal components are particular linear combinations of the p random variables $\{x_1, x_2, x_3, …, x_p\}$ with three important properties. The first one is the principal components are uncorrelated. The second one is the first principal component has the highest variance and the second principal component has the second highest variance and so on. The third one is the total variation in all the principal components combined equal to the total variation in the original variables $\{X_1, X_2, X_3, …, X_p\}$. The new variables with such properties are easily obtained from eigenanalysis of the covariance matrix or the correlation matrix of $\{X_1, X_2, X_3, …, X_p\}$ [14]. Let the original data X be a n x p data matrix of n observations on each of p variables $(X_1, X_2, …, X_p)$ and let R be a p x p sample correlation matrix of $X_1, X_2, …, X_p$. If $(\lambda_1, e_1), (\lambda_2, e_2), (\lambda_3, e_3), … (\lambda_p, e_p)$ are the p eigenvalue and eigenvector Pairs of the matrix R, $\lambda_1 \geq \lambda_2 \geq \lambda_3 … \geq \lambda_p \geq 0$, then ith sample principal component of an observation vector $x = (x_1, x_2, x_3, …, x_p)$ is

$$y_i = e_i\, z$$
$$y_i = e_{i1}z_1 + e_{i2} z_2 + e_{i3}z_3 + … + e_{ip} z_p , i = 1,2,3,.., p \qquad (3)$$

Where
$e_i = (e_{i1}, e_{i2}, e_{i3},…, e_{ip})$ is the ith eigenvector.
And
$Z = (z_1, z_2, z_3, …, z_p)$ is the vector of standardized observations defined as
$$z_k = x_k - \bar{x}_k , k=1, 2, 3, …, p \qquad (4)$$

Where $\bar{x}_k$ is the sample mean of the variable $x_k$. The ith principal component has sample variance $\lambda_i$ and the sample covariance or correlation of any pair of principal components is equal to zero. The PCA produces a set of independent variables so the total variance of a sample is the sum of all the variances accounted for by the principal components. The correlation between any two variables is

$$\rho_{i,j} = \frac{cov(x_i,x_j)}{\sigma_i \sigma_j} \qquad (5)$$

Where $\sigma_i$ is the standard deviation of $x_i$ which is a sample of data. The principal components of the sample correlation matrix have the same properties as principal components from a sample covariance matrix. As all principal components are uncorrelated, the total variance in all of the principal components is

$$\lambda_1 + \lambda_2 + \cdots + \lambda_p = p \qquad (6)$$

The principal components produced by the covariance matrix are different from the principal components produced by the correlation matrix. Eigenvalues have larger weights because of some values are much larger than others. Since The NSL-KDD data set has many items with varying scales and ranges so the correlation matrix will use.

### 3.3 Applying PCA to Outlier Detection

PCA applied as an outlier detection method. In applying PCA, there are two main issues, (1) how to interpret the set of principal components and (2) how to calculate the notion of distance. First, each eigenvalue of a principal component corresponds to the relative amount of variation it encompasses. The larger the eigenvalue is the more significant its corresponding projected eigenvector should be. Therefore, the most significant principal components sorted before the least significant principal components. If a new data item projected along the upper set of the significant principal components, it is likely that the data item can be classified without projecting along all of the principal components. Second, the data sample can represent by the axes of eigenvectors of the principal components. Those axes considering a normal when the data sample is the training set of normal network connections. If any points lie outside these axes by far distance then the data connection would exhibit abnormal data connection.

Outliers measured using the Mahalanobis distance are presumably network connections that are anomalous, any network connection with a distance greater than the threshold value (t) is considered an outlier. In this work, any outlier represents an attack. Consider the sample principal components of an $y_1, y_2, ..., y_p$ observation x where

$y_i = e_i\, z$ , i =1,2,... , p
$z_k = x_k - \bar{x}_k$ , k=1, 2, 3, ..., p

The sum of scores that are squares of the partial principal component is equal to the principal component score

$$\sum_{i=1}^{p} \frac{y_i^2}{\lambda_i} = \frac{y_1^2}{\lambda_1} + \frac{y_2^2}{\lambda_2} + \cdots + \frac{y_p^2}{\lambda_p} \qquad (7)$$

Equating to the Mahanobolis distance of the observation X from the mean of the normal sample data set [15].

Anomaly detections Needs an offline training or learning phase whether those methods are outlier detection, statistical models, or association rule mining. PCA has two clearly separate phases (the offline and online detection phases). These two separate phases are an advantage for hardware implementation. Another advantage of PCA is reduction of features. As we will show in our experiment, PCA effectively reduces the number of processed features from 41 to 10 or 6 features.

The outline steps involved in PCA are shown in (figure 1). Training data take as input and a mean vector of each sample calculate in the offline phase. Ideally, these data sets are a snapshot of activity connections in a real network environment. In addition, these data sets should contain only normal connections. Second, correlation matrixes calculate from the training data.

A correlation matrix normalizes all of the data by calculating the standard deviation. Next, eigenanalysis performed on the correlation matrix to create independent orthonormal eigenvalue and eigenvector pairs. The set of principal components can use in online analysis because of these pairs. Finally, the sets of principal components sort by eigenvalue in descending order. The eigenvalue is a relative measure of the variance of its corresponding eigenvectors.

Using dimensionality-reducing method such as PCA to extract the most significant principal components, so only a subset of the most important principal components needs to classify any new data. In addition to using the most significant principal components (q) to find intrusions, we have found that it is helpful to look for intrusions along a number of least-significant components (r) as well.

The major principal component score calculated by the most significant principal components and the minor principal component score calculated by the less significant principal components. Major principal component score (MajC) is used to detect severe deviations with large values of the original features. These observations follow the correlation structure of the sample data. Minor principal component score (MinC) is used to detect attacks may not follow the same correlation model. In this work, two thresholds needed to detect attacks. If the principal components sorted in descending order, then (q) is a subset of the highest values and is a subset of the smallest components. The MajC threshold is referred ($t_M$) while the MinC threshold is referred to ($t_m$). An observation (x) is an attack if

$$\sum_{i=1}^{q} \frac{y_i^2}{\lambda_i} > t_M \quad \text{Or} \quad \sum_{i=p-r+1}^{p} \frac{y_i^2}{\lambda_i} > t_m \tag{8}$$

The online portion takes major principal components and minor principal components and maps online data into the eigenspace of those principal components

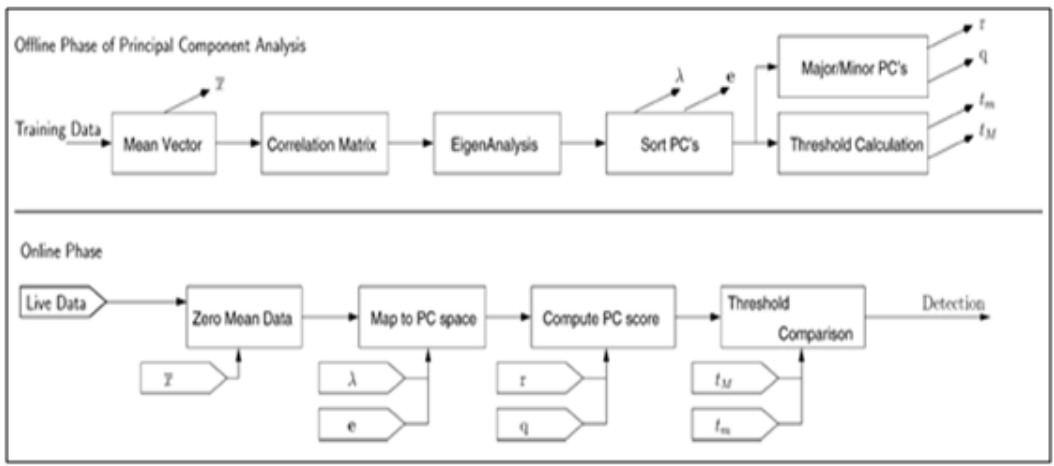

Figure (1) PCA For Network Intrusion Detection

## 4. EXPERIMENT

### 4.1 Data Set Description

Mostly all the experiments on intrusion detection are done on KDDCUP '99 dataset, which is a subset of the 1998 DARPA Intrusion Detection Evaluation data set and is processed extracting 41 features from the raw data of DARPA 98 data set. Defined higher level features that help in differentiating between "good" normal connections from "bad" attacks connections [16]. KDDCUP 99 data set can be used in host-based systems, network-based systems, signature systems and anomaly detection systems.

A connection is a sequence of Transmission Control Protocol (TCP) packets starting and ending with the time between which data come from a source IP address to a target IP address under some protocol. Each connection described as a normal or as an attack with defined the attack type. Each connection record consists of about 100 bytes [17].

KDD train and test set contains a huge number of records and huge number of redundant records. Almost about 78% and 75% of the records duplicated in the train and test set respectively. The classification will be wrong because of these redundant records and thus these records prevent classifying the other records that is not redundant. To solve this problem, a new dataset was developed NSL-KDD [18]. One copy of each repeated record was not removed in the KDD train and test set.

**4.2 Performance Measures**

Metrics, which are mainly used to evaluate the performance of classifiers are presented in [19], [20] and are given here for ready reference.

• The true positives (TP) are correct classifications and true negatives (TN) are correct classifications. True positive is the probability that there is an alert, when there is an intrusion.
• A false negative (FN) occurs when the outcome is incorrectly predicted as negative when it is actually positive.
• The true positive rate (TPR) is computed as

$$\text{TPR} = \frac{TP}{TP+FN} \quad (9)$$

• A false positive (FP) occurs when the outcome is incorrectly predicted positive when it is actually negative. The false positive rate computes as

$$\text{FPR} = \frac{FP}{FP+TN} \quad (10)$$

• Recall: The percentage of the total relevant documents in a database retrieved by your search computes as

$$\text{recall} = \frac{TP}{TP+FN} \quad (11)$$

• Precision: The percentage of relevant documents in relation to the number of documents retrieved is calculated as

$$\text{precision} = \frac{TP}{TP+FP} \quad (12)$$

• The overall success rate is the number of correct classifications divided by the total number of classifications is calculated as

$$\text{success rate} = \frac{TP+TN}{TP+TN+FP+FN} \quad (13)$$

$$\text{error rate} = 1 - \text{success rate} \quad (14)$$

**4.3 Experiment steps and results**

In our experiments we use KDDTrain_20Percent [21] in both the training and testing stages. The KDDTrain_20Percent contain 25192 connections records. The training data sets contain records of network connections labelled either as normal or as an attack. Each connection record made up of 41 different features related to the connection.

The 41 features are divided into three categories: basic features of TCP connections (1) □, content features of the connection (2)□, and traffic features (3)□ which are derived using a 2-s time window to monitor the relationships between connections. The same service and the same host information are included in The traffic-level features such as the number of connections in the past 2 s that have the same destination host as the current connection.

First, we select 6 features from the basic features of TCP connections which used with NIDS because these features do not need any host logs. Second, we add 4 features from traffic features, which based on time window and this collection (10 features), used in HIDS is shown in (Table 1).

Table 1 Feature Used in Our Experiment

| Feature name | Description | Type |
|---|---|---|
| Duration | Number of seconds of the connection | Continuous (1) |
| Protocol Type | Type of the protocol, e.g. tcp, udp, icmp . | Discrete (1) |
| Service | Network service on the destination, e.g., http, telnet, https, etc | Discrete (1) |
| Src-bytes | Number of data bytes from source to destination | Continuous (1) |
| Dst-bytes | Number of data bytes from destination to source | Continuous (1) |
| Flag | Normal or error status of the connection | Discrete (1) |
| Count | Number of connections from the same source as the current connection in the past two seconds | Continuous (3) |
| Sev-count | Number of connections to the same service as the current connection in the past two seconds from the same source | Continuous (3) |
| Dst-host-count | Number of connections to the same host as the current connection in the past two seconds | Continuous (3) |
| Dst-host-srv-count | Number of connections to the same service as the current connection in the past two seconds to the same host | Continuous (3) |

We used a Matlab program to design our IDS. Based on [22], we suggest using (q) major components that can explain about 50 - 70 percents of the total variation in the standardized features. When the original features are uncorrelated, each principal component from the correlation matrix has an eigenvalue equal to 1. So the minor components are those components whose variances or eigenvalues are less than 0.20, which would indicate some relationships among the features (r).

First step we selected 6 features and suggested using $q = 3$, $r = 0$. Second step we added 4 features and suggested using $q = 3$, $r = 2$. In a multiclass prediction, the result on a test set is often displayed as a two dimensional confusion matrix with a row and a column for each class.
Each matrix element shows the number of test examples for which the actual class is the row and the predicted class is the column. Good results correspond to large numbers down the main diagonal and small, ideally zero, off-diagonal elements. The confusion Matrix is showed on the (Table 2). The Performance Measures are shown in (Table 3) and (Table 4).

Table 2 Confusion Matrix

|  |  | Predicted Class | |
|---|---|---|---|
| **Actual Class** |  | Attack | Normal |
|  | Attack | TP | FN |
|  | Normal | FP | TN |

International Journal of Computer Networks & Communications (IJCNC) Vol.5, No.4, July 2013Table.3 Detection Attacks In All Steps

| Attacks | DOS | PROBE | R2l | U2r |
|---|---|---|---|---|
| Exist | 9234 | 2289 | 209 | 11 |
| Detection from step (1) | 8666 | 2212 | 28 | 1 |
| Detection from step (2) | 9028 | 2244 | 32 | 2 |

Table 4 metrics for all steps

| | Step (1) | | Step (2) | |
|---|---|---|---|---|
| **Metrics** | Normal class | Anomaly class | Normal class | Anomaly class |
| Recall and TPR | 0.9050 | 0.9288 | 0.7503 | 0.9628 |
| FPR | 0.0712 | 0.0949 | 0.0372 | 0.2496 |
| Precision | 0.9357 | 0.8952 | 0.9584 | 0.7719 |
| Overall success | 0.9161 | | 0.8493 | |
| Error | 0.0839 | | 0.1507 | |

Both recall and precision have good value in these two steps but one of steps can be used as NIDS another can be used as HIDS which has a better detection rate.

## 5. CONCLUSION AND FUTURE WORK

Future network intrusion detection system generation will most likely employ both signature detection and anomaly detection modules. Anomaly detection methods process a large amount of data in order to recognize anomalous behaviour or new attacks.

This paper used PCA as an effective way of outlier analysis. PCA is particularly useful because of its ability to reduce data dimensionality into a smaller set of independent variables from which new data can be classified.

This paper has two steps in its experiment. The first step takes six features from the basic features of TCP connections that can used in NIDS and this step has an overall success rate (0.9161) with high detection rate (0.9288). The second step takes ten features {six features from the basic features of TCP connections plus four features from traffic features} which can be used in HIDS and this step has an overall success rate (0.8493) with very high detection rate for Anomaly class (0.9628).

Plan for the future work is to use these two steps to make an integrated intrusion detection system by using relationship between these two steps.

155

**ACKNOWLEDGEMENTS**

Thanks to everyone who helped me in carrying out this work to the fullest

**REFERENCES**

[1]     Kumar, A., Maurya, H. C., Misra, R. (April 2013). A Research Paper on Hybrid Intrusion Detection System. International Journal of Engineering and Advanced Technology (IJEAT), volume-2, Issue-4, ISSN: 2249-895

[2]     Zargar, G. R. (October 2012). Category Based Intrusion Detection Using PCA. International Journal of Information Security, 3, 259-271.

[3]     Amparo, A. B., Noelia, S. M., Félix, M. C., Juan, A. S. and Beatriz, P. S. (25-27 April 2007). Classification of Computer Intrusions Using Functional Networks—a Comparative Study. Proceedings of European Symposium on Artificial Neural Networks (ESANN), Bruges. pp 579-584.

[4]     Ilgun, K., Kemmerer, R. A. and Porras, P. A. (1995). State Transition Analysis: A Rule-Based Intrusion Detection Approach. IEEE Transaction on Software Engineering, Vol. 21, No. 3, pp. 181-199.

[5]     Guyon, I. and Elisseff, A. (2003). An Introduction to Variable and Feature Selection. Journal of Machine Learning Research, Vol. 3, pp. 1157-1182.

[6]     Chou, T. S. Yen, K. K. and Luo, J. (2008). Network Intrusion Detection Design Using Feature Selection of Soft Computing Paradigms. International Journal of Computational Intelligence, Vol. 4, No. 3, pp. 196-208.

[7]     Chakraborty, B. (2005). Feature Subset Selection by Neuro-Rough Hybridization. Lecture Notes in Computer Science (LNCS), Springer, Heidelberg.

[8]     Sung, A. H. and Mukkamala, S. (2003). Identifying Important Features for Intrusion Detection Using Support Vector Machines and Neural Networks. Proceedings of International Symposium on Applications and the Internet (SAINT) pp. 209-216.

[9]     Chebrolu, S. Abraham, A. and Thomas, J. (2005). Feature Deduction and Ensemble Design of Intrusion Detection Systems. Computers and Security, Elsevier Science, Vol. 24, No. 4, pp. 295-307.

[10]    Abraham, A. and Jain, R. (2004). Soft Computing Models for Network Intrusion Detection systems, Springer, Heidelberg.

[11]    Abraham, A. Grosan, C. and Vide, C. M. (2007) "Evolutionary Design of Intrusion Detection Programs," International Journal of Network Security, Vol. 4, No.3, pp. 328-339.

[12]    Gyanchandani, M. Yadav, R. N. Rana, J. L. (December 2010). Intrusion Detection using C4.5: Performance Enhancement by Classifier Combination. International Journal on Signal and Image Processing, Vol. 1, No. 03

[13]    Boutsidis, C. Mahoney, M. W. and Drineas, P. (2008). Unsupervised Feature Selection for Principal Components Analysis. Proceedings of the 14th ACM Sigkdd International Conference on Knowledge Discovery and Data Mining, Las Vegas, pp. 61-69

[14]    Jolliffe, I. T. (2002). Principal component analysis. $2^{nd}$ Ed. Springer, Verlag, NY.

[15]    Jobson, J. D. (1992). Applied Multivariate Data Analysis, Volume II: Categorical and Multivariate Methods. New York: Springer Verlag.

[16]    Stolfo, J. Fan, W. Lee, W. Prodromidis, A. and Chan, P.K. (2000). Cost-based modeling and evaluation for data mining with application to fraud and intrusion detection. DARPA Information Survivability Conference.

[17]    The KDD Archive. KDD99 cup dataset, 1999: http://kdd.ics.uci.edu/databases/kddcup99/kddcup99.html

[18]    Tavallaee, M. Bagheri, E. Lu, W. and Ghorbani, A. (2009). A Detailed Analysis of the KDD CUP 99 Data Set. Proceedings of the Second IEEE Symposium on Computational Intelligence for Security and Defense Applications (CISDA).

[19]    Srinivasulu, P. Nagaraju, D. Ramesh Kumar, P. and Nagerwara Rao, K. (June 2009). Classifying the Network Intrusion Attacks using Data Mining Classification Methods and their Performance Comparison. International Journal of Computer Science and Network Security, Vol.9 No.6, pp 11-18.

[20]    Shyu, M. Chen, S. Sarinnapakorn, K. and Chang, L. (2003). A novel anomaly detection scheme based on principal component classifier. Proceedings of the IEEE foundation and New Directions

International Journal of Computer Networks & Communications (IJCNC) Vol.5, No.4, July 2013
of Data Mining Workshop, in conjunction with the Third IEEE International Conference on Data Mining (ICDM03), pp. 172-179
[21]   The NSL-KDD Data set: http://nsl.cs.unb.ca/NSL-KDD/
[22]   Shyu, M. Chen, S. Sarinnapakorn, K. Chang, L. (2003). A Novel Anomaly Detection Scheme Based on Principal Component Classifier. IEEE Foundations and New Directions of Data Mining Workshop, in conjunction with ICDM'03, pp. 171-179.

## AUTHORS

**Mohamed Faisal**    received the B.sc degree from Assiut University (in 2010). After working as a Network security engineer (from 2011) in information network at Sohag University and Research Assistant in the Department of Electrical Engineering, at Sohag University (from 2011), He has been a demonstrator in MUST University (since2012). He finished his Preliminary Master in June 2012 in the Department of Electrical Engineering, at Assuit University.

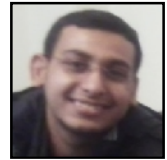

**Tarik Kamal**    received the B.sc. and M.sc. degrees, from Assuit University in 1975 and 1980, respectively. He received the Dr. Eng. degree from France in 1986. After working as a demonstrator (from1975) and as an assistant lecturer (from 1981), He has been a lecturer in the Department of Electrical Engineering at Assuit University since 1987. His research interest includes signal processing, image processing and communication network. He is a supervisor of Information network at Assiut University.

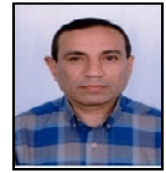

**Abdel-Fattah Mahmoud**    received the B.sc. and M.sc. degrees, from Assuit University in 1976 and 1981, respectively. He received the Dr. Eng. degree from Maryland University in 1990. After working as demonstrator (from1978), Assistant Lecturer (from 1981) in Assuit University, Visitor Professor of Department of Mechanical Engineering, University of Texas, United States of America (from September 1991 to August 1993), associate professor (from 1995) in Assuit University, Visitor Professor of the Department of Electrical Engineering, Kanazawa University, Japan, (from April 1996 to April 1997) and Visitor Professor of the University Technology in Malaysia (from February 2006 - March 2006), he has been a professor in the Department of Electrical Engineering, Assuit University since 2000. He has been a dean of Engineering College, Assuit University since 2011.

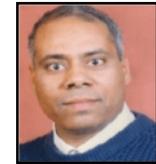